\documentclass[aps,showpacs,eqsecnum,twocolumn,superscriptaddress,usegraphicx]{revtex4}

\usepackage{amsmath}
\usepackage{latexsym}
\usepackage{graphicx}
\usepackage{psfig}

\begin{document}


\title{Influence of self-gravity on the runaway instability of black hole-torus systems}



\author{Pedro J. Montero}
\affiliation{Max-Planck-Institute f{\"u}r Astrophysik,
Karl-Schwarzschild-Str. 1, 81748, Garching bei M{\"u}nchen, Germany}

\author{Jos\'{e} A. Font}
\affiliation{Departamento de Astronom\'{\i}a y Astrof\'{\i}sica,
Universidad de Valencia, Dr. Moliner 50, 46100 Burjassot, Spain}

\author{Masaru Shibata}
\affiliation{Yukawa Institute of Theoretical Physics, Kyoto
  University, Kyoto 606-8502, Japan}

\date{\today}

\begin{abstract}
Results   from  the   first  fully   general   relativistic  numerical
simulations  in  axisymmetry  of  a  system formed  by  a  black  hole
surrounded by  a self-gravitating torus in  equilibrium are presented,
aiming to assess the influence  of the torus self-gravity on the onset
of the  runaway instability. We  consider several models  with varying
torus-to-black  hole  mass  ratio  and angular  momentum  distribution
orbiting in equilibrium around a non-rotating black hole. The tori are
perturbed  to induce  the mass  transfer towards  the black  hole. Our
numerical simulations show that  all models exhibit a persistent phase
of  axisymmetric  oscillations  around  their equilibria  for  several
dynamical   timescales   without  the   appearance   of  the   runaway
instability, indicating  that the self-gravity  of the torus  does not
play a critical role favoring the onset of the instability, at least
during the first few dynamical timescales.
\end{abstract}


\pacs{
04.25.Dm, 
04.70.Bw, 
95.30.Lz 
}


\maketitle

{\it Introduction.}  Self-gravitating  tori orbiting black holes (BHs)
may form  after the merger  of a  binary system formed  by a BH  and a
neutron   star  (NS)   or   the   system  formed   by   two  NS   (see
e.g.\cite{Rezzolla10a} and references  therein). In addition, they may
also be the result of  the gravitational collapse of the rotating core
of   massive  stars   \cite{Woosley93,Paczynski98}.   State-of-the-art
numerical simulations  have started to  provide quantitative estimates
of     the      viability     of     such      systems     to     form
\cite{Rezzolla10a,Baiotti08a,Duez09,Etienne09,Shibata06b,Shibata05,Shibata06a,Shibata06,Shibata07b}. 
As such  BH-torus  systems are  thought  to  be  the central  engine  for
gamma-ray   bursts (GRBs)    \cite{Piran99,Meszaros06},   understanding   its
formation, dynamics and stability properties is of high relevance.

In  particular,  the so-called  runaway  instability,  first found  by
Abramowicz, Calvani and Nobili \cite{Abramovici83}, is an axisymmetric
instability that could destroy  the torus on dynamical timescales.  In
a marginally stable torus, the  radial pressure gradient may drive the
transfer of  mass towards the BH  through the cusp-like  inner edge of
the torus.  Due to the accretion of mass and angular momentum, both the 
mass of  the BH and its  spin increase, and the gravitational  field changes, 
leading to  two possible evolutions: (i) if the cusp moves inwards  towards 
the BH, the mass transfer slows down and the  system is stable, or (ii) if the  
cusp moves deeper into the  torus, mass accretion  will increase,  and the  
accretion process will be runaway unstable.

The  numerical  study of  the  runaway  instability  has so  far  been
investigated       under      different       approximations      (see
e.g.~\cite{Font02a,Daigne04}).  Abramowicz et al.~\cite{Abramovici83},
assuming  a pseudo-Newtonian  potential for  the BH,  constant angular
momentum distribution in the torus and an approximate treatment of the
disk's  self-gravity, found  that the  instability occurs  for  a wide
range of  initial models.  More  detailed studies based  on stationary
models,   either  assuming  a   pseudo-Newtonian  potential   for  the
BH~\cite{pseudo}        or       being        fully       relativistic
calculations~\cite{nishida96}, indicated that  the self-gravity of the
disk  favors the  instability, by  arguing that,  as a  result  of the
accretion process,  the cusp  would move closer  to the center  of the
torus  than  in  non   self-gravitating  disks.   However,  there  are
additional  parameters  which  have  a  stabilizing  effect:  (i)  the
rotation of the BH~\cite{wilson}, and (ii) the most important one, the
radial distribution of specific  angular momentum, increasing with the
radial distance~\cite{power-law}.

\begin{table*}
\begin{center}
\caption{Main properties of the equilibrium models studied in units of
  $c=G=M_{\rm \odot}=1$  (unless shown otherwise). From  left to right
  the   columns  show:   the   type  of   specific  angular   momentum
  distribution,  the  torus-to-BH  mass  ratio, the  position  of  the
  maximum density point  $r_{\rm max}$, the position of  the inner and
  outer radii of the torus $r_{\rm in}$ and $r_{\rm out}$, the maximum
  rest-mass density and the orbital  period at the center of the torus
  $t_{\rm orb}$.}

\label{tab1}
\begin{tabular}{lccccccccc}
\hline Model &$j-{\rm law}$ &$M_{\rm t}/M_{\rm BH}$ & $r_{\rm
  max}$ & $r_{\rm in}$ & $r_{\rm out}$ &$\rho_{\rm max}$ $(g/cm^{3})$ &
$t_{\rm orb}$ \\

\hline
M1     & const     & 0.1 & 7.17  & 4.92  & 10.17 & $3.189 \times10^{14}$ & 147.81 \\
M2     & const     & 1.0 & 8.87  & 4.02  & 19.97 & $2.202 \times10^{14}$ & 199.54 \\
M3     & non-const & 0.1 & 10.47 & 4.92  & 19.97 & $3.902 \times10^{13}$ & 245.37 \\
M4     & non-const & 0.5 & 10.02 & 4.07  & 19.97 & $1.538 \times10^{14}$ & 229.91 

\end{tabular}
\end{center}
\end{table*}

The  first time-dependent,  general relativistic  hydrodynamical (GRH)
axisymmetric simulations of the runaway instability were performed by  Font and Daigne~\cite{Font02a,Daigne04}. The BH
evolution was assumed to follow a sequence of stationary BH spacetimes
of increasing  mass and angular  momentum, controlled by the  mass and
angular   momentum  transferred   from  the   torus,  whose
self-gravity  was  neglected.   The first  work~\cite{Font02a},  which
focused  on  tori  with  constant  distribution  of  specific  angular
momentum  $l\equiv -u_{\varphi}/u_{t}$,  with $u_{\varphi}$  and $u_t$
being the corresponding components of the 4-velocity $u_{\mu}$, showed
that the system  is runaway unstable on a  dynamical timescale. On the other  
hand, the  second work~\cite{Daigne04}  showed that thick  disks with
non-constant  specific angular momentum distributions, increasing outwards  
with the radial distance according to a  power law $l=Kr^{\alpha}$, are stable
for very  small values  of the angular  momentum slope  $\alpha$ (much
smaller  than  the   Keplerian  value  $\alpha=0.5$),  confirming  the
prediction of stationary studies.

Despite the progress that has  been made the existing works are still
not conclusive, mainly due to
the absence of  important physics in the modeling.   The complexity of
handling the  presence of a  spacetime singularity in addition  to the
hydrodynamics and  the self-gravity of the accretion  torus, make full
GRH  simulations of  such  systems very  challenging. The  simulations
presented in this {\it Letter}  accomplish the goal of assembling this
important physics to provide a conclusive answer about the likelihood
of the instability on the first few dynamical timescales, which cause
the most concern in the context of models of short GRB. Although we do not exclude the onset of the instability on
much longer timescales, it is only meaningful in connection with other
relevant instabilities, specially the magnetorotational instability.

{\it Numerical setup.} The numerical simulations have been carried out
with the {\tt nada} code (see~\cite{nada} for details and tests) which
solves the Einstein equations (using the BSSN approach, the ``cartoon"
method, and the moving puncture approach) coupled to the GRH equations
(solved with a high-resolution  shock-capturing scheme based on 3rd-order
PPM  reconstruction  and the  HLLE  approximate  Riemann solver).   In
addition, we use a standard  $\Gamma$-law equation of state (EOS), for
which the pressure is expressed as a function of the rest-mass density
and  specific  internal  energy as  $P=\rho\epsilon(\Gamma-1)$,  where
$\Gamma$ is the adiabatic exponent.  For the vacuum region outside the
torus  we use  a  dynamically unimportant  artificial atmosphere  with
rest-mass  density $\rho_{\rm  thr}\sim  10^{-8}\rho_{\rm max}$.   The
evolution equations are  integrated by the method of  lines, for which
we use  a 4th-order Runge-Kutta  scheme. For the  simulations reported
here we use  an equidistantly spaced $(x,z)$ grid  with a grid spacing
$\Delta x=\Delta z=0.05M_{\rm BH}$  and $N_{x}\times N_{z}=600\times 600$ points
to cover a computational domain, $0\leq x \leq L$ and $0 \leq z \leq L
$,  with  $L=30M_{\rm BH}$.  Unless  otherwise  stated  we  use units  in  which
$c=G=M_{\rm \odot}=1$.

{\it Initial data.} Compact  equilibrium configurations for a BH-torus
system  are obtained  in the  moving puncture  framework (we  refer to
\cite{Shibata07} for details). We  adopt a $\Gamma=4/3$ polytropic EOS
to  mimic   a  degenerate  relativistic  electron   gas,  and  construct initial
configurations  using  either constant  or non-constant
specific  angular  momentum  distributions,  defined as  $j  \equiv  h
u_{\varphi}$ ($h$ being  the specific enthalpy). A list  of the models
along  with  their main  features  is  given  in Table~\ref{tab1}.  We
consider four different tori around a non-rotating BH (of mass $M_{\rm
  BH}= M_{\odot}$).   Following   \cite{Font02a,Daigne04,Zanotti03},  the   EOS
polytropic constant $\kappa$ is  chosen such that the torus-to-BH mass
ratio,  $M_{\rm t}/M_{\rm BH}$,  is 0.1,  0.5 or  1, depending  on the
model. We  note that existing  simulations of NS/NS and  BH/NS mergers
\cite{Rezzolla10a,Baiotti08a,Shibata06a,Shibata06b,Shibata09}   yield  
$M_{\rm  t} \sim  0.01-0.2$  M$_{\rm  \odot}$.  As these  configurations  are  not
overflowing the cusp, we introduce an initial perturbation to induce a
small  mass  transfer   through  the  inner  edge  of   the  tori  (as
in~\cite{Zanotti03}).  In our simulations  we simply perturb the $v^x$
component of  the 3-velocity  of the torus  as $v^{x}  \approx -\eta$,
which otherwise would initially be  zero. For each model the numerical
simulations are  stopped at $t\sim  2000$, which corresponds  to $\sim
10$ ms (between 8 to 10 orbital periods depending on the model), since
at late times the growth of the Hamiltonian constraint violation would
lead to a loss of  accuracy for the spacetime evolution. Nevertheless,
the time  scale we are considering  in our simulations  would allow to
identify the  runaway instability, if  present, as it could  even take
place within one orbital period for  the more massive models M2 and M4
(see~\cite{Font02a,Daigne04}).

\begin{figure*}
\includegraphics[angle=0,width=7.5cm]{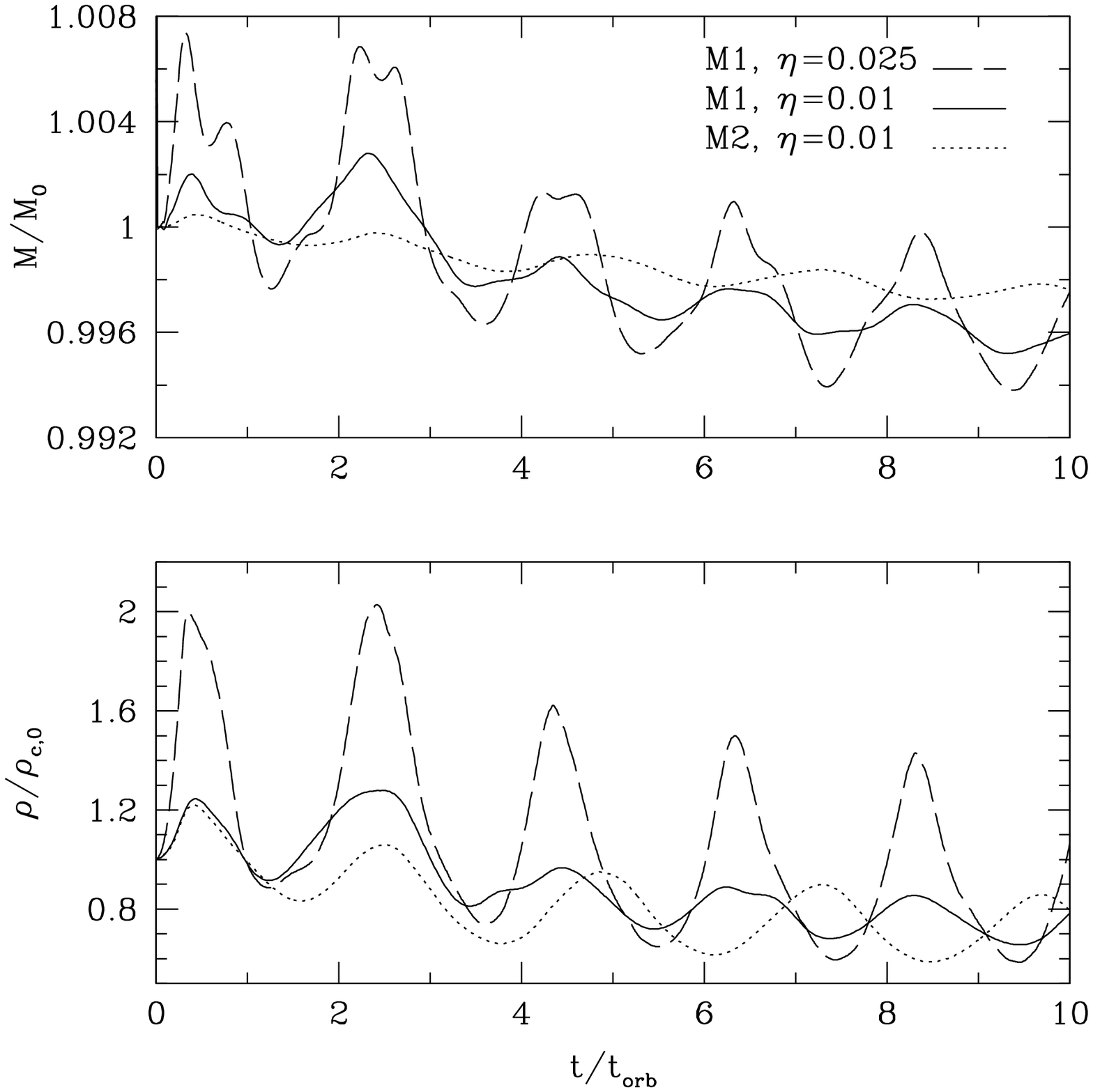}
\includegraphics[angle=0,width=7.5cm]{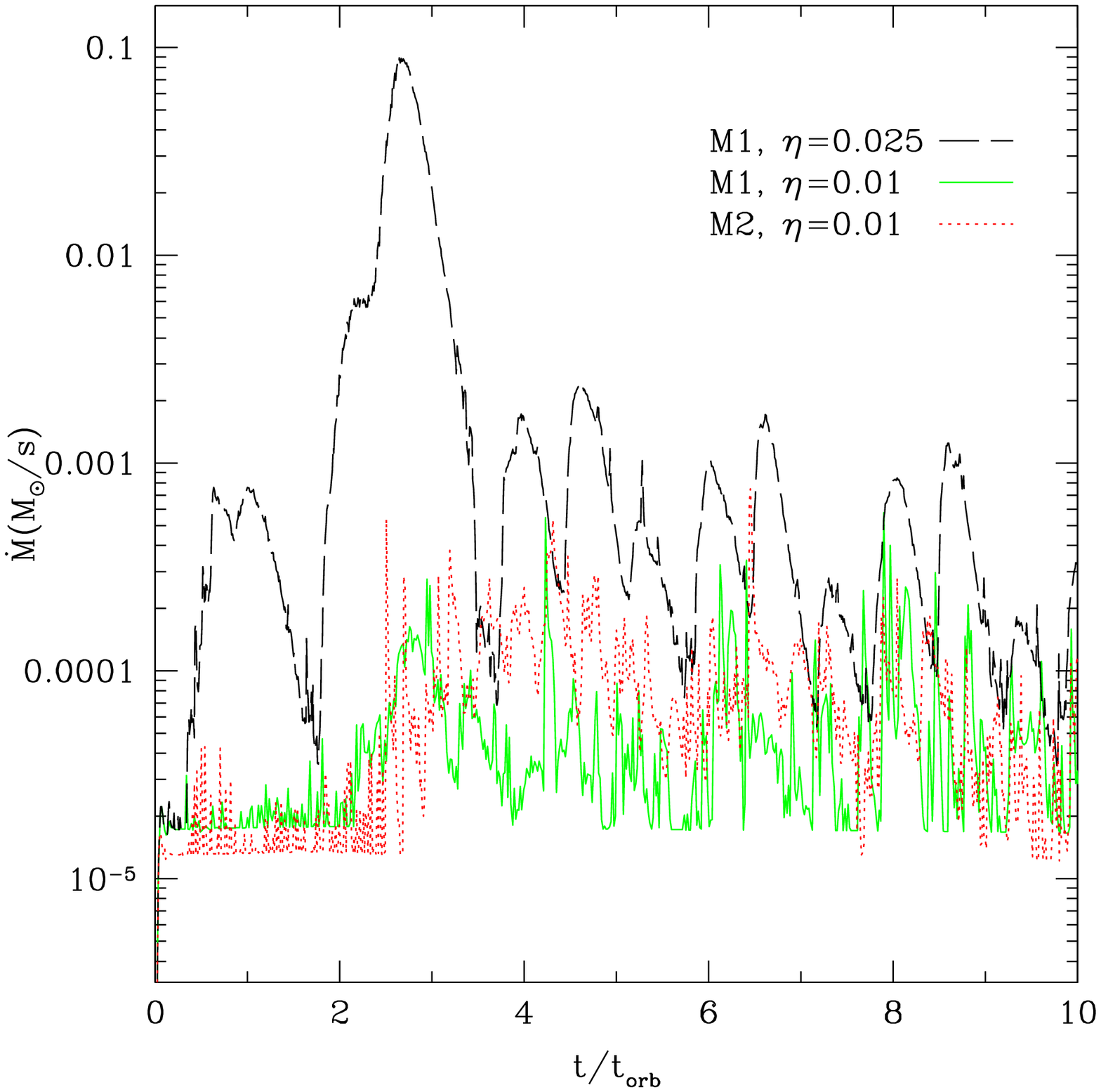}
\caption{Left panel:  Time evolution of the total  rest-mass (top) and
  central rest-mass  density, each of  them normalized to  its initial
  value,  for the evolution  of models  M1 and  M2. Right  panel: Mass
  accretion rate evolution for models M1 and M2.}
\label{fig1y2}
\end{figure*}

{\it Results.} The left panel in Figure~\ref{fig1y2} displays the time
evolution of  the total rest-mass and central  rest-mass density, each
of them normalized  to its initial value, for  the evolution of models
M1(solid and  dashed lines) and M2  (dotted line). M1 is  a model with
$j$-constant  and with  an initial  rest-mass of  $M_{\rm t}=0.1M_{\rm
  BH}$,  thus representing  a  model with  torus-to-BH  mass ratio  in
agreement  with results  obtained  by simulations  of  NS/NS or  BH/NS
mergers~\cite{Rezzolla10a,Baiotti08a,Shibata06a,Shibata06b,Shibata09}.   
For  this initial model, we have considered two different initial perturbations,
$\eta=0.01$ (solid  line) and  $\eta=0.025$ (dashed line)  to evaluate
its  influence on  the  overall dynamics.   As  expected, the  initial
perturbation  triggers a  phase  of axisymmetric  oscillations of  the
torus  around  its  equilibrium   which  are  present  throughout  the
simulation. Such oscillations induce a small outflow of matter through
the cusp towards the BH.  This, however, does not reduce significantly
the total rest-mass  of the torus, plotted in the  upper panel. At the
end of the simulation $(t\sim 10 \, t_{\rm orb})$ the rest-mass of the
torus M1  is conserved up to  about $1\%$. Therefore, the  BH mass and
spin do not  increase with time significantly, and  the torus shows no
sign of the runaway instability. Further information about the process
of accretion is obtained  from the right panel in Figure~\ref{fig1y2},
which shows the  time evolution of the mass  accretion rate. We notice
that the  mass accretion rate  for model M1  is larger the  larger the
initial perturbation  is. For $\eta=0.025$, there is  an initial stage
of very small mass transfer through  the inner edge of the torus which
lasts for  about half  an orbit  $(t\sim 80)$. This  is followed  by a
stage  in which the  oscillatory behavior  of the  rest-mass accretion
rate, signature of the induced oscillations, is obvious throughout the
numerical evolution. Interestingly,  during the oscillation phase, the
mass  flux does  not increase  in amplitude  with time,  as  one would
expect   prior    to   the   onset   of    the   runaway   instability
(see~\cite{Font02a,Daigne04}). Instead, it  reaches a maximum of about
$\dot{M} \sim 0.1$ M$_{\rm  \odot}$/s after the second orbital period.
Later  it decreases  and exhibits  a series  of oscillations  around a
lower value,  never showing any  signature of exponential  growth. The
mass accretion rates of the  perturbed tori are in good agreement with
the  maximum  expected   accretion  rates  for  hyper-accreting  disks
associated with the central engine of gamma-ray bursts, which could be
as high as  $\dot{M} \sim 0.01-1$ M$_{\rm \odot}$/s,  depending on the
formation mechanism \cite {Chen}. In the  case of a disk formed by the
merger of a NS with another compact object (either a NS or a BH), most
of the  material in  the accretion  disk would be  accreted on  a time
scale  comparable to  the viscous  time scale  $t_{\rm  visc}\sim 0.1$
s.  In the  collapsar scenario,  such  high accretion  rates could  be
sustained  for $\sim  10\,$ s  as the  disk is  fed by  the in-falling
stellar material.

We consider next  a significantly more massive torus,  model M2, while
keeping the same  rotation law.  The total rest-mass  of this torus is
$M_{\rm  t}=1.0 M_{\rm  BH}$.  We  use a  small  initial perturbation,
$\eta=0.01$,  because due to  the larger  coordinate dimension  of the
torus,  a larger  perturbation would  cause  its outer  parts to  move
outside  the computational  domain.   Despite being  more massive  the
overall dynamics of M2 is very  similar to the one discussed above for
the less  massive torus,  M1. As expected,  the time evolution  of the
central  rest-mass density,  displayed  in the  left  panel of  Figure
\ref{fig1y2}  with  a dotted  line,  shows  a  series of  axisymmetric
oscillations during the  entire length of the simulation.   As we have
observed for the low mass torus, the amplitude in the evolution of the
mass flux does  not increase with time and does not  lead to the onset
of the instability.
\begin{figure}
\includegraphics[angle=0,width=7.5cm]{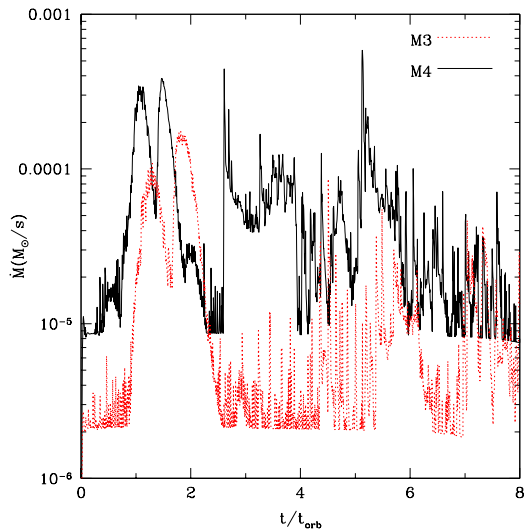}
\caption{Mass accretion rate for models M3 and M4.}
\label{fig2y3}
\end{figure}
We  note  that these  results  do  not  actually contradict  results
obtained for  non-self-gravitating tori with  constant distribution of
angular momentum~\cite{Font02a}.   Notice that despite  the difference
in the  definition of the specific angular  momentum, the $j$-constant
condition in  our models leads to  $l$-constant up to  a difference of
the order  of $10^{-5}$.  In  addition, models of  ~\cite{Font02a} and
the ones  considered here satisfy the  condition $j_{\rm torus}>j_{\rm
  ISCO}$  throughout the  evolution, where  ISCO stands  for innermost
stable  circular  orbit. However,  those  previous studies  considered
initial models  which were  overflowing the  cusp in  order to
induce  a  large  stationary  accretion  rate,  which  varied  between
$\dot{M} \sim  0.1-34.0$ M$_{\rm \odot}$/s.  Unstable runaway behavior
in  the case of  mass accretion  rates of  $\dot{M} \sim  0.1$ M$_{\rm
  \odot}$/s were  found on  a timescale of  $100$ $t_{\rm orb  }$, and
only  on a  dynamical timescale  for the  largest values  of  the mass
flux. Similar  results were  found by \cite{Zanotti03}  introducing an
initial  perturbation  on the  equilibrium  tori.  Since  in our  case
$\dot{M}\sim   10^{-3}-10^{-4}$  M$_{\rm   \odot}$/s   throughout  the
simulations, the  exponential growth of  the mass accretion  would not
manifest itself even  on such timescales. This is in agreement with the 
recent work of~\cite{Rezzolla10a} where a systematic study of  the 
BH-torus systems produced  by the merger of  unequal-mass NS  binaries 
was presented, concluding that self-gravitating tori
with mass accretion rates as high as $\dot{M} \sim 2.0$ M$_{\rm \odot}$/s
were stable on the dynamical timescales investigated.

Motivated by the influence of  different rotation laws on the onset of
the  instability~\cite{Font02a}  we have  also  carried out  numerical
evolutions  of  two  $j$-non-constant  models,  M3 and  M4,  with  two
different torus-to-BH  mass ratios, 0.1 and  0.5 respectively. Despite
the difference in  the rotation law with respect to  models M1 and M2,
the dynamics  is very similar, as  inferred from the  evolution of the
mass accretion  rate displayed in  Figure~\ref{fig2y3}. No exponential
growth  of  the  mass  flux   is  found  and,  therefore,  no  runaway
instability is present.

Studies  \cite{pseudo,nishida96} mainly  based  on stationary  models,
either  assuming  a  pseudo-Newtonian  potential  for  the  BH  and  a
Newtonian potential for the self-gravity  of the torus, or being fully
relativistic calculations indicated that  the self-gravity of the disk
favors the  instability. The  simulations presented here  overcome the
limitations  of preceding  works.  Our  results indicate  that  in the
general case in  which mass accretion rates are  consistent with those
expected for hyper-accreting disks,  and in which the angular momentum
distribution   increases   with  the radial   distance,   the  effect   of
self-gravity is not sufficient to lead to unstable accretion.

 Despite the simplifying assumptions of our models (magnetic fields or
 detailed microphysics are not considered), our results are consistent
 with   the   expected  dynamics   and   mass   accretion  rates   for
 hyper-accreting disks  and do not  challenge the time  scale required
 for producing a GRB. Moreover, although magnetic fields have not been
 considered in our simulations, it is  likely that these do not play a
 critical role in the context  of the runaway instability. Despite the
 fact that  numerical simulations of  magnetized NS mergers  are still
 scarce and the results  are inconclusive, simulations by \cite{Liu08}
 indicate that the effect of the magnetic fields during the merger and
 the initial phase  of the BH-torus lifetime  is not dramatic (as
 reflected by the similarities  between the gravitational waves in the
 magnetized  and unmagnetized  cases),  rather these  are expected  to
 become more  important during the secular evolution  of the accretion
 torus.

{\it  Conclusions.} We  have presented  results from  the  first fully
general relativistic numerical simulations  in axisymmetry of a system
formed  by a  BH  surrounded by  a marginally-stable  self-gravitating
torus aiming to assess the  influence of the torus self-gravity on the
onset  of  the  runaway  instability. Several  models  with  different
torus-to-BH mass  ratio and  angular momentum distributions  have been
considered.   The tori  have been  perturbed to  induce  mass transfer
towards the BH. Our numerical simulations show that all models exhibit
a   persistent  phase  of   axisymmetric  oscillations   around  their
equilibria for several dynamical  timescales without the appearance of
the runaway instability. Thus, the  self-gravity of the torus does not
play a critical role favoring  the onset of the instability.  Clearly,
to  investigate additional  $m=1$ non-axisymmetric  features  that may
play  a  role  on  the  dynamics  of the  system  3D  simulations  are
required. Nevertheless,  the robustness of our  results in axisymmetry
on  the  influence of  self-gravity  on  the  runaway instability  are
confirmed by  simulations we have also performed  with the independent
2D code  of~\cite{Shibata03}. These simulations  and the investigation
of non-axisymmetric instabilities  of self-gravitating tori around BHs
will be presented elsewhere.

{\it Acknowledgments.}   We thank Frederic Daigne,  Ewald M\"uller and
Yudai  Suwa for  useful  comments.   This work  was  supported by  the
Collaborative Research Center  {\it Gravitational Wave Astronomy} of
the  Deutsche  Forschungsgesellschaft   (DFG  SFB/Transregio  7),  the
Spanish   {\it    Ministerio   de   Educaci\'on    y   Ciencia}   (AYA
2007-67626-C03-01)  and   by  Grant-in-Aid  for   Scientific  Research
(21340051) and  by Grant-in-Aid for Scientific  Research on Innovative
Area (20105004) of the Japanese Monbukagakusho.


\bibliographystyle{apsrev}


\end{document}